\documentclass[osajnl,twocolumn,showpacs,superscriptaddress,10pt]{revtex4-1}

\usepackage{amsmath,amssymb,graphicx}
\usepackage{latexsym} 
\usepackage{amsmath,amsthm}
\usepackage{ifpdf}
\usepackage{epstopdf}
\usepackage{dcolumn}
\usepackage{bm}
\usepackage{braket}
\usepackage{color}

\begin{document}

\def \beq{\begin{equation}}
\def \eeq{\end{equation}}
\def \bea{\begin{eqnarray}}
\def \eea{\end{eqnarray}}
\def \bem{\begin{displaymath}}
\def \eem{\end{displaymath}}
\def \P{\Psi}
\def \Pd{|\Psi(\boldsymbol{r})|}
\def \Pds{|\Psi^{\ast}(\boldsymbol{r})|}
\def \Po{\overline{\Psi}}
\def \bs{\boldsymbol}
\def \bl{\bar{\boldsymbol{l}}}

  \ifpdf
    \else
    \fi

\title{Generalized Uncertainty Principle and Analogue of Quantum Gravity in Optics}

\author{Maria Chiara Braidotti}
		\email{mariachiara.braidotti@isc.cnr.it}
    \affiliation{Institute for Complex Systems, National Research Council (ISC-CNR), Via dei Taurini 19, 00185 Rome (IT).}
		\affiliation{Department of Physical and Chemical Sciences, University of L'Aquila, Via Vetoio 10, I-67010 L'Aquila (IT).}
\author{Ziad H. Musslimani}
		\affiliation{Department of Mathematics, Florida State University, Tallahassee, Florida 32306-4510, USA}
\author{Claudio Conti}
    \homepage{http://www.complexlight.org}
		\affiliation{Institute for Complex Systems, National Research Council (ISC-CNR), Via dei Taurini 19, 00185 Rome (IT).}
    \affiliation{Department of Physics, University Sapienza, Piazzale Aldo Moro 5, 00185 Rome (IT).}
    

\begin{abstract}
The design of optical systems capable of processing and manipulating ultra-short pulses and ultra-focused beams is highly challenging with far reaching fundamental technological applications. One key obstacle routinely encountered while implementing sub-wavelength optical schemes is how to overcome the limitations set by standard Fourier optics. A strategy to overcome these difficulties is to utilize the concept of {\it generalized uncertainty principle} (G-UP) that has been originally developed to study quantum gravity. In this paper we propose to use the concept of G-UP within the framework of optics to show that the generalized 
Schr\"{o}dinger equation describing short pulses and ultra-focused beams predicts the existence of a minimal spatial or temporal scale which in turn implies the existence of maximally localized states. Using a Gaussian wavepacket with complex phase, we derive the corresponding generalized uncertainty relation and its maximally localized states. We numerically show that the presence of nonlinearity helps the system to reach its maximal localization. Our results may trigger further theoretical and experimental tests for practical applications and analogues of fundamental physical theories.
\end{abstract}
\maketitle
\section{Introduction}
For a given optical system such as a fiber or an imaging apparatus, understanding the shortest achievable pulse or the thinnest producible spot is an issue of paramount importance for a large number of practical applications and fundamental reasons. 
In this regard, Fourier optics is the reference paradigm for designing ultrafast temporal processes, and imaging systems \cite{Born80}. In Fourier optics the uncertainty principle relates the spectral content of a beam to its spatial size thus allowing one to engineer optical systems and their numerical aperture for specific applications. However, the formalism of Fourier optics cannot be used for beams with size comparable to their wavelength because of the onset of nonparaxial effects.\\

Recent developments in the area of super resolved microscopy \cite{Klar:99}, involve light beams with size much smaller than the wavelength in which case the standard Heisenberg uncertainty principle (H-UP) breaks down. See\-min\-gly in the temporal domain, the uncertainty principle intervenes in deter\-mi\-ning the minimal duration for transform limited pulses \cite{AgrawalBook}. However for ultra-short pulses \cite{AgrawalBook} higher holder dispersion forbids to predict the shortest accessible signal with simple Fourier optics. \\
To generalize the uncertainty principle to tackle the challenge of determining the smallest possible beam or the shortest optical pulse for a given spatial and temporal dispersion, there is the need of looking at novel techniques.
In the following we show that un\-ex\-pec\-tedly quantum gravity furnishes a possible road.\\
Many quantum gravity models predict a space discretization which results in having a minimal uncertainty length $\Delta x_{min}$. This feature is inferred by a modification of the standard uncertainty principle of quantum mechanics to a generalized uncertainty principle which in the simplest form can be written as 
\begin{equation}
\Delta x\Delta P> \frac{\hbar}{2}\left[1+\beta(\Delta P)^2\right] \;,
\end{equation}
where $\Delta P$ is the momentum uncertainty and $\beta > 0$ is a parameter that takes into account the deviation from the standard Heisenberg uncertainty principle. The possible validity of a G-UP has been studied for decades as the key to solve fundamental problems in physics as the transplanckian problem of the Hawking radiation, the modification of the blackbody radiation spectrum, corrections to cosmological constants and to the black-hole entropy \cite{Chang02,Trabelsi10}.\\
Despite all these investigations, the value of $\beta$ is unknown and its particular expression in terms of other physical constants, such as, the Planck length, varies depending on the various quantum gravity theories. It is often expressed in terms of the dimensionless parameter $\beta_0=M_P^2 c^2 \beta$, with $M_P$ being the Planck mass, and $c$ is the speed of light in vacuum.
Letting $G$ denote the gravi\-ta\-tional constant,  and $M_P=\sqrt{\hbar c /G}$ the Planck mass, $\beta_0$ is also written as  
\begin{equation}
\beta_0=\frac{\hbar c^3}{G}\beta \;.
\label{beta0def}
\end{equation}
Some authors affirm that $\beta_0\cong 1$, but a recent ana\-ly\-sis po\-ses the limit $\beta_0<10^{34}$ \cite{Das08,Das09}.
Even in the case $\beta_0\cong 10^{34}$, accessing experimentally measurable effects of a G-UP appears to be prohibitively difficult. In this regard, finding analogues is hence very important either to test the new reported G-UP predictions or to provide insights for further theoretical developments and novel experiments.  \\
There is an unexpected ``link" between quantum gravity and nonparaxial and ultrafast 
optics \cite{Conti2014}. The key point is that the first order non-paraxial theory (and seemingly the theory of pulse propagation with higher order dispersion) is formally identical to the modified quantum Schr\"odinger equation that is studied in the G-UP literature \cite{Das08}: 
\begin{equation}
i\hbar\partial_t\psi=\frac{\hat p^2}{2m}\psi+\frac{\beta}{3m}\hat p^4\psi, 
\label{GSE}
\end{equation} 
with $\hat p=-i\hbar\partial_x$ being the quantum momentum, $\psi$ is the quantum wave-function and $m$ the particle mass.
This mathematical analogy allows one to describe and test nonparaxial and ultrafast regimes for optical propagation in terms of the paradigms developed in the G-UP framework. As we detail below, in the optical analogs the values of $\beta$ are such that we can foreseen doable emulations of the physics at the Planck scale. \\
In this paper, we develop the concept of {\it generalized uncertainty principle} (G-UP) in the framework of li\-near and nonlinear optics. The generalized linear and nonlinear 
Schr\"{o}dinger equation describing short pulses and ultra-focused beams is used to
predict the existence of a minimal spatial or temporal scale. As a result, maximally localized states exist and their properties are discussed. The theoretical results are tested for a Gaussian wavepacket with complex phase. An explicit inequality for the generalized uncertainty relation is derived along with its corresponding maximally localized modes. We numerically show that the presence of nonlinearity helps the system to reach its maximally localized state.\\\\
The manuscript is organized as follows: in section II, we propose the higher order propagation equation and show that it is formally equivalent to the generalized 
quantum Shr\"{o}dinger equation (\ref{GSE}) both in the temporal and spatial domain. We derive the explicit expression of the $\beta$ parameter in our optical analogue. In section III, we find the expression of the G-UP for optics, deriving the minimal uncertainty length $\Delta x_{min}$, and analyze its properties in the case of a chirped Gaussian wavepacket. In section IV, we introduce and evaluate the Maximally Localized States, which are the states which satisfy the G-UP strictly. As a final part, in section V, we show that these maximally localized states naturally occur in the nonlinear regime. Conclusions are drawn in section VI.
\section{Higher order Schr\"odinger equation}
\subsection{Spatial case and nonparaxiality}
We start this section by showing how the wave equation can be formally ``mapped" to the quantum Schr\"odinger equation (\ref{GSE}). To this end, we consider a unidimensional Helmholtz equation for the electric field $\mathcal{E}$ and propagation direction $z$
\begin{equation}
\partial^2_z \mathcal{E}+\partial_x^2 \mathcal{E}+k_0^2 \mathcal{E}=0\;,
\label{helm1}
\end{equation}
where $k_0=2\pi/\lambda$ with $\lambda$ being the wavelength. We remark that vectorial effects are not present in vacuum \cite{Ciattoni:05,Longhi09_analog,Aiello2005}. Equation (\ref{helm1}) admits forward and backward pro\-pa\-gating waves with longitudinal (i.e., in the $z-$direction) wavenumber
\begin{equation}
\label{dr}
k_z=\pm\sqrt{k_0^2-k^2} \;,
\end{equation}
with $k$ being the transverse wavenumber. Retaining only forward propagating beams, 
the forward projected Helm\-holtz equation (FPHE) reads \cite{Kolesik2004a}
\begin{equation}
 i \partial_z \mathcal{E}+\sqrt{\partial_x^2+k_0^2}\mathcal{E}=0\text{.}
\label{fwdhelm}
\end{equation}
In general, the dispersion relation (\ref{dr}) describes both spatially periodic as well as evanescent waves. However, in this paper, we shall consider dynamics of narrowly localized beams (in momentum space) corresponding to Fourier mode $k$ satisfying the condition $|k| \ll k_0$. With this in mind, we expand the dispersion relation (\ref{dr}) in powers of $k^2$
and obtain (retaining terms up to order $k^4$) the first-order non-paraxial equation \cite{Lax75}
\begin{equation}
i\partial_z \mathcal{A} =-\frac{1}{2k_0}\partial_x^2 \mathcal{A}+\frac{1}{8 k_0^3}\partial_x^4 \mathcal{A}  \;,
\label{npe}
\end{equation}
with $\mathcal{A}=\mathcal{E}e^{-ik_0z}$.
To further establish the connection between G-UP in quantum mechanics and its optical 
analog, we identify the value of the parameters $\beta$ and $\beta_0$. Letting $\partial_x=-\frac{i \hat p}{\hbar}$ and $z=ct$ one obtains the following expression for the $\beta$ parameter \cite{Conti2014}
\begin{equation}
\beta=\frac{3}{8}\left(\frac{\lambda}{h}\right)^2\text{.}
\label{betaKMM}
\end{equation}
The formal identity between the unidirectional FPHE and the SE allows one to provide an expression for the parameter $\beta$ shown in Eq.(\ref{betaKMM}), and hence of its corre\-spon\-ding normalized $\beta_0$. In the optical case, from Eq.(\ref{beta0def}) and (\ref{betaKMM}), $\beta_0$ can be written as:
\begin{equation}
\beta_0=\frac{3}{8}\frac{M_P^2}{m^2}=\frac{3}{8}\frac{c^3(\lambda/2\pi)^2}{G \hbar}\text{.}
\label{contieq1}
\end{equation}
We report in Table (\ref{table1}) values of $\beta_0$ obtained from Eq.(\ref{contieq1}). In \cite{Das08} it has been estimated $\beta_0<10^{34}$.
We hence observe that, in the optical analogue, G-UP effects for the photon are expected to be much more pronounced being $\beta_0~=~10^{55}$.

\begin{table}[h]
\centering
     \begin{tabular}{ | l || l | l | l |}
     \hline
           & $\lambda$(m) & m(kg)    &  $\beta_0$  \\ \hline \hline
     photon       & $10^{-6}$    & $10^{-36}$  &  $10^{55}$                       \\ \hline
     $\gamma$ ray & $10^{-12}$   & $10^{-31}$ & $10^{45}$                       \\ \hline
     neutron      & $10^{-15}$   & $10^{-27}$ & $10^{39}$                       \\\hline
     \end{tabular}
     \caption{$\beta_0$ calculated from Eq. (\ref{contieq1}), for the neutron with $v\cong c$. Note that $\beta_0\simeq 1$ in the quantum gravity literature.}
\label{table1}
 \end{table}
Quantum gravity effects are often considered to be un-observables, even if some possibilities have been reported in the li\-te\-ra\-ture \cite{Das08,Das09} but also questioned \cite{ACamelia13}. In our analogue, one can see that nonparaxial regimes for light allows to test some concepts introduced in the G-UP li\-te\-ra\-ture. In the same perspective, mathematical tools developed in the G-UP framework furnish novel roads for nonparaxial and ultrafast light propagation.
\subsection{Temporal case}

The formal analogy found in the spatial case can be also extended to the temporal domain for which the temporal dynamics of a highly dispersive pulses is governed by \cite{AgrawalBook}
\begin{equation}
i\frac{\partial A}{\partial z}-\frac{\beta_2}{2}\frac{\partial^2 A}{\partial t^2}-\frac{\beta_4}{4!}\frac{\partial^4 A}{\partial t^4}=0 \;.
\label{temporal}
\end{equation}
We consider the case of  dispersion-flattened fiber with zero third order dispersion ($\beta_3=0$). \cite{AgrawalBook}
By defining the following rescaled variables $z = Tc$ and $t=\frac{X}{c}$ where $T$ and $X$ represent the new time and space variables, one finds
\begin{eqnarray}
\beta=-\frac{\beta_4 c^2}{8\hbar^2\beta_2} \;, \\
\beta_0=-\frac{\beta_4 c^5}{8G\hbar\beta_2} \;.
\end{eqnarray}
Since the parameters $\beta$ and $\beta_0$ are positive definite, we have the 
constraint $\beta_2\beta_4<0$. Typical values for the parameters $\beta$ and $\beta_0$ can be obtained by considering an optical fiber with dispersion coefficients $\beta_2=0.49$ $\mbox{ps}^2/\mbox{Km}$ and $\beta_4~=~-1.1~\times~10^{-7}~\mbox{ps}^4/\mbox{m}$ \cite{Droques:13}
which gives 
$$\beta\simeq 10^{56} \frac{\mbox{s}^2}{\mbox{Kg}^2\mbox{m}^2} \;,\;\;\;\;\;
\beta_0\simeq 10^{57}.$$
\\
As detailed in the following, G-UP predicts a maximal localization corresponding in the temporal case to a mi\-ni\-mum time uncertainty $\Delta T_{min}=\hbar\sqrt{\beta}/c$. For $\beta\simeq1$, $\Delta T_{min}\simeq\hbar/c \simeq 10^{-31}$s which should give the maximal temporal resolution. For the values of $\beta$ obtained in our analogy and given above, $\Delta T_{min}\simeq 10^{-15} $s.
This means that maximally localized states of quantum gravity correspond to pulses of duration of the order of femto\-se\-conds and demonstrates that laboratory emulations of the physics at the Planck scale are indeed accessible.
\section{Optical G-UP: A unified framework}
We stress that G-UP is typically assumed as a postulate in modern quantum gravity theories. Our goal here is to show that the G-UP formalism is also relevant for spatial and temporal optical wave propagation. We hence follow a different strategy and derive the gene\-ra\-lized uncertainty relation starting from the governing dynamical evolution equation. Thus, the starting point is the normalized higher order propagation equation
\begin{equation}
i\partial_z\psi+\frac{1}{2}\partial_x^2\psi-\frac{\varepsilon}{8}\partial_x^4\psi=0 \;,
\label{NORMALIZED}
\end{equation}
where $\psi$ is the envelop wave-function proportional to the electric field, $z$ is the propagation direction, $x$ re\-pre\-sents either the spatial or temporal variable and
 $\varepsilon$ is a dimensionless parameter. In the spatial case $\varepsilon=1/k_0 Z_d$, where $Z_d$ is the diffraction length. 
In the temporal case $\varepsilon=-\beta_4/(3\beta_2 T_0^2)$, with $T_0$ being the initial temporal pulse duration.\\
Throughout the rest of the paper the forward Fourier transform is defined by
\begin{equation}
\label{FT-1}
F(f)=\tilde{f}(k) = \frac{1}{\sqrt{2\pi}} \int_{-\infty}^\infty dx f(x)e^{-ik x}\;,
\end{equation}
with the inverse given by
\begin{equation}
\label{IFT-1}
f(x) = F^{-1}(\hat{f})=\frac{1}{\sqrt{2\pi}} \int_{-\infty}^\infty dk \tilde{f}(k)e^{ik x} \;.
\end{equation}
The first step in obtaining the generalized uncertainty relation is to define the generalized momentum $K$. Letting $\varepsilon>0$, we take the Fourier transform of Eq.~(\ref{NORMALIZED}) and obtain
\begin{equation}
\label{eq1}
i\partial_z\tilde{\psi}-\left(\frac{k^2}{2}+\varepsilon\frac{k^4}{8}\right)\tilde{\psi}=0\;,
\end{equation}  
where $\tilde{\psi}$ is the Fourier transform of $\psi.$ Defining $K^2 \equiv k^2+\varepsilon k^4/4$, Eq.~(\ref{eq1}) then takes the equivalent form
\begin{equation}
i\partial_z\tilde{\psi}-\frac{K^2}{2}\tilde{\psi}=0 \;,
\end{equation}  
with generalized momentum $K$ approximately given by $K \approx k+\frac{\varepsilon}{8}k^3$, for $\varepsilon\ll 1$ or if the band is limited. In this regard, the inverted dispersion relation reads
\begin{equation}
k\simeq K-\frac{\varepsilon}{8}K^3.
\label{inv}
\end{equation}
We remark that the $\varepsilon$ expansion has limited the values of the transverse wavevector $k$. This in turn would set certain limits on the accessible values for the generalized momentum $K$ as well. We underline that in the $K$-space the scalar product takes the form
\begin{equation}
\langle\tilde{\psi}(K)|\tilde{\phi}(K)\rangle = \int_{-\infty}^\infty
\frac{\tilde{\psi}^*(K)\tilde{\phi}(K)}{1+\beta K^2}dK.
\label{scalar_product}
\end{equation}
In order to derive the desired uncertainty principle, we first recall the Heisenberg-Robertson inequality \cite{Robertson29}: for two operators $\hat{A}$ and $\hat{B}$ with uncertainty 
$\Delta A$ and $\Delta B$ we have
\begin{equation}
\Delta A \Delta B \geq \frac{1}{2}\left|\langle\left[\hat A,\hat B\right]\rangle\right|.
\end{equation} 
\begin{figure}[t]
	\centering
		\includegraphics[scale=0.4]{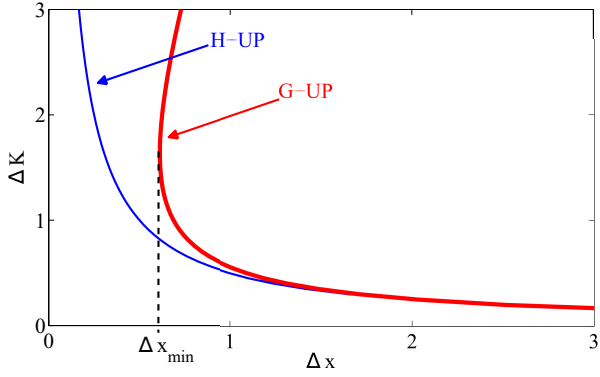}
	\centering
	\caption{Heisenberg uncertainty principle (thin line) and its generalization (thick line). Note that, in the generalized case, the increasing of $\Delta K$ does not imply a reduction in $\Delta x$ and a minimum $\Delta x_{min}$ exist. In the picture, we used a large value of $\varepsilon$ $(\varepsilon=1)$ to emphasize the differences between the lines.}
\label{fig:gup}
\end{figure}
The generalized uncertainty principle can be obtained by using the commutation rule: 
\begin{equation}
\left[\hat x,f(\hat k)\right]=i\frac{\partial f(\hat k)}{\partial k}.
\end{equation} 
With this at hand, we have the following result
\begin{equation}
\label{comm}
[\hat x,\hat K(\hat k)] = i\left(1+ 3\varepsilon \hat k^2/8\right).
\end{equation}
Substituting the expression for $k$ (see Eq.~(\ref{inv})) in Eq.~(\ref{comm}) and 
keeping terms up to order $\varepsilon$ we find
\begin{equation}
[\hat x,\hat K(\hat k)]=i\left(1+ 3\varepsilon \hat K^2/8\right)\;,
\end{equation}
from which we obtain
\begin{equation}
\label{ineq}
\Delta x \Delta K \geq \frac{1}{2}\left(1+\frac{3}{8}\varepsilon \langle\hat K^2\rangle\right).
\end{equation} 
If one assumes $\langle\hat K\rangle=0,$ the inequality (\ref{ineq}) reduces to
\begin{equation}
\Delta x \Delta K \geq \frac{1}{2}\left(1+\frac{3}{8}\varepsilon \Delta K^2\right).
\label{dxdk_gen}
\end{equation} 
This is the generalized uncertainty principle associated with Eq.(\ref{NORMALIZED}) given in dimensionless form. Figure~\ref{fig:gup} shows a graphical representation of Eq.~(\ref{dxdk_gen}). An important aspect related to Eq.~(\ref{dxdk_gen}) is the existence of a minimal position uncertainty 
\begin{equation}
\Delta x_{{\rm min}}=\sqrt{\frac{3\varepsilon}{8}}.
\label{dx0}
\end{equation}
We remark that this is valid for $\epsilon\ll 1$ or for a li\-mi\-ted bandwidth. This theory predicts maximally loca\-li\-zed states, which are the ones that satisfy strictly the generalized uncertainty principle and hence have a width equals to $\Delta x_{{\rm min}}$. On the other hand we can see that this theory agrees with the Heisenberg uncertainty principle. Indeed, for $\varepsilon=0$ and at a fixed $\Delta K$, it is possible to focus a given beam 
until $\Delta x=\frac{1}{2}\frac{1}{\Delta K}$, and then $\Delta x_{{\rm min}}\rightarrow 0$ as $\Delta K \rightarrow \infty$.
\subsection{Gaussian wave-packets and minimal uncertainty}
\begin{figure}[h!]
	\centering
		\includegraphics[scale=0.4]{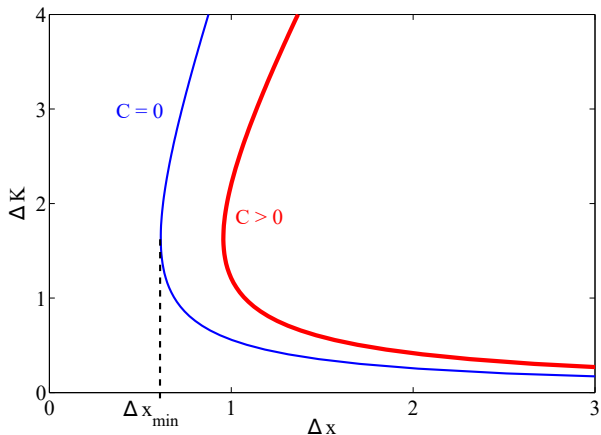}
	\caption{Generalized uncertainty principle in the presence of a chirp $C$ $(\varepsilon=1)$. Thick line denote the G-UP for $C=1$.}
\label{fig:gup_C}
\end{figure}

In this section we apply the results obtained so far for a chirped Gaussian beam \cite{AgrawalBook} 
by calculating its uncertainty relation $\Delta x \Delta K$.  We assume a wave-function in the form
\begin{equation}
\psi=\frac{1}{\sqrt{\pi}x_0}e^{-\frac{x^2}{2x_0^2}(1+iC)}
\label{gaussian_beam}
\end{equation}
where $C$ is a chirp parameter (tilt in the spatial case). Its corresponding form in momentum space is given by
\begin{equation} 
\tilde{\psi} = \frac{\sqrt{x_0}}{\sqrt[4]{\pi}}\frac{1}{1+iC}e^{-\frac{k^2 x_0^2}{2(1+iC)}}.
\end{equation}
Straightforward calculations show
\begin{equation} 
\Delta x^2=\int x^2|\psi|^2 dx=\frac{1}{2}x_0^2
\end{equation}
\begin{equation} 
\Delta k^2=\int k^2|\tilde{\psi}|^2 dk=\frac{1}{2}\frac{1+C^2}{x_0^2}
\end{equation}
and uncertainty relation \cite{AgrawalBook}
\begin{equation} 
\Delta x\Delta k=\frac{1}{2}\sqrt{1+C^2}.
\label{dxdkC}
\end{equation}
In this case, the minimal beam waist is 
\begin{equation}
\Delta x_{min}=\frac{x_0}{\sqrt{2}\sqrt{1+C^2}}.
\label{dxmin_c}
\end{equation}
$\Delta x_{min} \rightarrow 0$ for $C\rightarrow\infty$.\\
We proceed taking the generalized momentum as $K=k+\frac{\varepsilon}{8}k^3$, and hence at the lowest order in $K$ we have
\begin{equation} 
\begin{aligned}
\Delta K^2&=\int K^2|\psi(K)|^2 \frac{dK}{1+\frac{3}{8}\varepsilon K^2}=\\
&=\int dk \left(k+\frac{\varepsilon}{8}k^3\right)^2|\tilde{\psi}(k)|^2 .
\end{aligned}
\end{equation}
Expanding the above integral at first order in $\varepsilon$
\begin{equation}
\begin{aligned}
&\int \left(k+\frac{3}{8}\varepsilon k^2\right)^2|\tilde{\psi}(k)|^2dk=\\
&\approx\frac{1}{2}\frac{(1+C^2)}{x_0^2}\left(1+\frac{3}{8}\varepsilon\frac{1+C^2}{x_0^2}\right).
\end{aligned}
\end{equation}
Thus, the generalized uncertainty principle reads as
\begin{equation}
\begin{aligned}
\Delta x\Delta K&=\frac{1}{2}\sqrt{1+C^2}\sqrt{1+\frac{3}{8}\varepsilon\frac{1+C^2}{x_0^2}}=\\
&\simeq\frac{1}{2}\sqrt{1+C^2}(1+\frac{3}{8}\varepsilon\Delta K^2)\;,
\label{dxdk_gauss}
\end{aligned}
\end{equation}
to leading order in $\varepsilon$ (see Fig.~\ref{fig:gup_C}). The relation (\ref{dxdk_gauss}) matches the general uncertainty principle (\ref{dxdk_gen}) for $\varepsilon\neq 0$. Moreover if $\varepsilon=0$ and the chirp goes to zero, we obtain the standard Heisenberg relation from Eq. (\ref{dxdk_gauss}). From Eq. (\ref{dxdk_gauss}), we compute the minimal value of $\Delta x$:
\begin{equation}
\Delta x_{min}=\sqrt{\frac{3}{8}\varepsilon(1+C^2)}+o(\varepsilon^2)
\end{equation}
where, for $C\rightarrow\infty$, $\Delta x_{min} \rightarrow \infty$. This means that there is a minimal value of $\Delta x_{min}$ which is the one found previously in Eq.(\ref{dx0}) [see Table (\ref{table2})].
\begin{table}
\centering
     \begin{tabular}{ | c || c | c |}
     \hline
																		& H-UP & G-UP     \\ \hline \hline
     $\Delta x$       							& $0$    			& $\Delta x_{min}$   \\ \hline
     $\Delta x$ for $C\rightarrow\infty$ & $0$ 				& $\infty$  \\ \hline
     \end{tabular}
     \caption{Minimal $\Delta x$ for standard and generalized uncertainty principle, with a chirp parameter $C$.}
\label{table2}
 \end{table}

\subsection{The generalized position operator}
In standard quantum mechanics, eigenstates of the position operator $\hat x$, corresponding to ideally localized wave-functions with $\Delta x=0$, form a basis of the Hilbert space. In the G-UP literature, states with $\Delta x=0$ are not physically acceptable as the position operator is not self-adjoint. In this framework one considers the {\it maximally localized states}, that satisfy Eq.(\ref{dx0}), i.e., $\Delta x=\Delta x_{min}$, as \emph{quasi}-position eigenstates. In our analogy, these states correspond to the mostly localized beams (within the adopted first order non\-pa\-ra\-xial approximation) or to the shortest light pulses one can achieve in the presence of second and forth order dispersion.   \\
Our intent here is to derive their expression with the re\-fe\-rence to our normalized model Eq.(\ref{NORMALIZED}). We follow the treatment reported in \cite{Kempf95}.
We start from the eigenstates of the generalized momentum operator $\hat{K}$.
\begin{eqnarray}
\hat K\psi(K)&=&K\psi(K)\\
\hat x\psi(K)&=&i\left(1+\frac{3}{8}\varepsilon K^2\right)\partial_K\psi(K)
\end{eqnarray}
where the $\hat x$ representation in the $K$ basis is $i(1+\frac{3}{8} \varepsilon K^2)\partial_K$, as one can verify by
\begin{equation}
\begin{aligned}
\left[\hat x,\hat K\right]\psi(K)&=\left(\hat x\hat K-\hat K\hat x\right)\psi(K)=\\\nonumber
&=i\left(1+\frac{3}{8}\varepsilon K^2\right)\psi(K),
\end{aligned}
\end{equation}
which gives the commutation relation found previously.
The operator $\hat x$ and $\hat K$ are symmetric, that is
\begin{eqnarray}
\left(\bra{\psi}\hat K\right)\ket{\phi}&=&\bra{\psi}\left(\hat K\ket{\phi}\right)\\
\left(\bra{\psi}\hat x\right)\ket{\phi}&=&\bra{\psi}\left(\hat x\ket{\phi}\right),
\end{eqnarray}
with respect to Eq. (\ref{scalar_product}) and the following completeness and orthogonality relations hold:
\begin{eqnarray}
1=\int_{-\infty}^{+\infty}\frac{dK}{1+\beta K^2}\ket{K}\bra{K}\\
\langle K|K'\rangle=\left(1+\beta K^2\right)\delta(K-K').
\end{eqnarray}
In order to compute eigenstates of the $\hat x$ operator, we consider the $\hat x$ eigenvalue equation in the $K$ space
\begin{equation}
\hat x \psi(K)=\lambda\psi(K),
\end{equation}
where we can write $\hat x$ explicitly as $i(1+\beta K^2)\partial_K$ so 
\begin{equation}
i\left(1+\beta K^2\right)\partial_K \psi(K)=\lambda\psi(K).
\end{equation}
Solving this equation, we find the normalized position eigenfunction in the $K$ space
\begin{equation}
\psi(K)=\sqrt{\frac{\sqrt{\beta}}{\pi}}e^{-ì\frac{\lambda}{i\sqrt{\beta}}\mbox{arctan}\left(K\sqrt{\beta}\right)}.
\label{mls}
\end{equation}
Equation (\ref{mls}) is still normalizable for $\lambda \in \mathbb{C}$ and  in the general case we have 
\begin{equation}
\psi(K)=\sqrt{\frac{Im(\lambda)}{\mbox{sinh}\left[\frac{\pi Im(\lambda)}{\sqrt{\beta}}\right]}}e^{-ì\frac{\lambda}{i\sqrt{\beta}}\mbox{arctan}\left(K\sqrt{\beta}\right)}.
\label{psi_di_k}
\end{equation}
For Im$(\lambda)\rightarrow 0$ we obtain Eq.(\ref{mls}). We remark that $\lambda$ can be a complex number as $\hat x$ is not self-adjoint.

\section{Evaluation of the Maximally Localized States}
A maximally localized state $\psi^{ML}_{\xi}$ is defined by:
\begin{eqnarray}
\langle\psi^{ML}_{\xi}|\hat x|\psi^{ML}_{\xi}\rangle=\xi\\
(\Delta x)_{\psi^{ML}_{\xi}}=\Delta x_{min}.
\end{eqnarray}
Following Heisenberg \cite{Heisenberg1927}, we start from
\begin{equation}
\left\| \left(x-\langle x\rangle+\frac{\langle[x,K]\rangle}{2(\Delta K)^2}(K-\langle K\rangle) \right)|\psi\rangle \right\|\geq 0,
\end{equation}
which implies that 
\begin{equation}
\Delta x\Delta K\geq\frac{|\langle[x,K]\rangle|}{2}.
\end{equation}
If a state $|\psi\rangle$ satisfies $\Delta x\Delta K=|\langle[x,K]\rangle|/2$, we have
\begin{equation}
\left(x-\langle x\rangle+\frac{\langle[x,K]\rangle}{2(\Delta K)^2}(K-\langle K\rangle) \right)|\psi\rangle=0.
\label{posnorm}
\end{equation}
Therefore equation (\ref{posnorm}) is used to find the MLS. Hereafter, we use the following notation:
\begin{eqnarray}
\langle x\rangle&=&\xi\\
\langle K\rangle&=&K_1\\
\langle K^2\rangle&=&K_2\\
\left[x,K\right]&=&i(1+\beta K^2).
\end{eqnarray}

From equation (\ref{posnorm}) we have in the $K$ space
\begin{equation}
i\partial_K\psi=\frac{-i}{1+\beta K^2}\left[\xi-\frac{i(1+\beta K_2^2)}{2(\Delta K)^2}(K-K_1)\right]\psi.
\end{equation}
Solving by variable separation, $\psi(K)$ reads as
\begin{widetext}
\begin{equation}
\psi(K)=\psi(0)\mbox{exp}\left[\left(-\frac{i\xi}{\sqrt{\beta}}+\frac{(1+\beta K_2^2)}{2(\Delta K)^2}\frac{K_1}{\sqrt{\beta}}\right)\mbox{arctg}(\sqrt{\beta}K)\right]\times\left[1+\beta K^2\right]^{-\frac{1+\beta K_2^2}{2(\Delta K)^2}\frac{1}{2\beta}}.
\end{equation}
\end{widetext}
For $\langle K\rangle=0$ and $\Delta K=\sqrt{K_2}=1/\sqrt{\beta}$
\begin{equation}
\psi^{ML}(K)=\psi(0)\frac{e^{-\frac{i\xi}{\sqrt{\beta}}\mbox{arctg}(\sqrt{\beta}K)}}{(1+\beta K^2)^{1/2}}
\label{mls_K}
\end{equation}
where $\xi \in\mathbb{R}$. \\
Imposing the normalization $\Braket{\psi^{ML}|\psi^{ML}}=1$ we have $\psi(0)=\sqrt{2\sqrt{\beta}/\pi}$. \\
From equation (\ref{mls_K}), one can verify that 
\begin{equation}
\begin{aligned}
(\Delta K)^2&=\frac{1}{\beta}\\
(\Delta x)^2&=\beta,
\end{aligned}
\end{equation}
for $\langle x\rangle=\xi=0$.\\
One can also verify that these states have finite energy $\langle \hat H\rangle=K^2=1/\beta$.\\
These states are not mutually orthogonal, i.e.
\begin{equation} 
\langle \psi_{\xi'}^{ML}(K')|\psi_{\xi}^{ML}(K)\rangle\neq\delta_{\xi',\xi}(K'-K).
\end{equation}
Indeed
\begin{equation} 
\begin{aligned}
\langle \psi_{\xi'}^{ML}|\psi_{\xi}^{ML}\rangle&=\int \frac{dK}{(1+\beta K^2)^2}\frac{2\sqrt{\beta}}{\pi}e^{-i(\xi-\xi')\frac{\mbox{arctg}(\sqrt{\beta}K)}{\sqrt{\beta}}}=\\
&=\frac{1}{\pi}\left[\frac{\xi-\xi'}{\sqrt{\beta}}-\left(\frac{\xi-\xi'}{\sqrt{\beta}}\right)^3\right]^{-1}\mbox{sin}\left(\frac{\xi-\xi'}{\sqrt{\beta}}\pi\right),
\label{psipsi}
\end{aligned}
\end{equation}
so they do not furnish a classical basis as in ordinary quantum mechanics. However they can be used as a re\-pre\-sen\-tation for wave-functions. Projecting a generic state $|\phi\rangle$ on $|\phi^{ML}\rangle$ we have:
\begin{widetext}
\begin{equation}
\phi(\xi)=\langle\psi_{\xi}^{ML}|\phi\rangle=\int_{-\infty}^{+\infty}dK\frac{2\sqrt{\beta}}{\pi(1+\beta K^2)^{3/2}}e^{i\xi\frac{\mbox{arctg}(\sqrt{\beta}K)}{\sqrt{\beta}}}\phi(K).
\label{newFT}
\end{equation}
\end{widetext}
In standard quantum mechanics this would correspond to the usual Fourier transform. Notably, this generalized Fourier transform is also invertible, as follows
\begin{equation}
\phi(k)=\int_{-\infty}^{+\infty}d\xi\frac{1}{\sqrt{8\pi\sqrt{\beta}}}(1+\beta K^2)^{1/2}e^{-i\xi\frac{\mbox{arctg}(\sqrt{\beta}K)}{\sqrt{\beta}}}\phi(\xi).
\label{inv_newFT}
\end{equation}
\begin{figure}[t]
	\centering
		\includegraphics[scale=0.40]{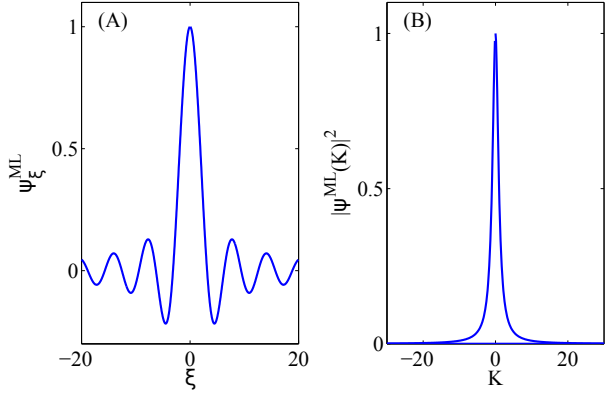}
	\centering
	\caption{(A) Maximally localized state $\langle\psi_{\xi}|\psi_0\rangle$ in the $\xi$-space; (B) Square modulus of the generalized Fourier transform of the maximally localized state.}
\label{fig:psix}
\end{figure}
In figure \ref{fig:psix} we show the characteristic profile of a maxi\-mally localized state defined by $\langle \psi_{\xi}|\psi_0\rangle$ and its gene\-ra\-lized Fourier transform. We remark the presence of the typical oscillations present in the maximally loca\-lized field.  

\section{Generalized Uncertainty Principle and nonlinearity}
In this section we show the way nonlinearity triggers the generation of maximally localized states. For that purpose, we consider nonlinear Schr\"odinger equation (NLS) with nonlocal nonlinearity and higher order diffraction \cite{MusslimaniPRA2014,MusslimaniPhysicaD2015}:
\begin{eqnarray}
\label{nls1}
i\frac{\partial\psi}{\partial z} = &-&\frac{1}{2}\frac{\partial^2\psi}{\partial x^2} +\frac{\varepsilon}{8} \frac{\partial^4\psi}{\partial x^4} +
\nonumber \\
&-& g \psi  \int_{-\infty}^{+\infty} G(x-x') |\psi (x')|^2 dx' \;,
\end{eqnarray}
where $g>0$ measures the strength of the nonlinearity; $\epsilon>0$ is the higher order diffraction coefficient and $G(x)$ is a kernel given by
\begin{eqnarray}
\label{kernel}
G(x)=\frac{e^{-|x|/\sigma}}{2\sigma} \;,
\end{eqnarray}
where $\sigma >0$ is a constant that characterizes the degree of nonlocality. Bound states for Eq.~(\ref{nls1}) are sought of in the form $\psi(x,z) = \phi(x) \exp (i \mu z)$ with $\phi$  satisfying the boundary value problem
\begin{eqnarray}
\label{nls2}
-\mu \phi &=& -\frac{1}{2}\frac{\partial^2\phi}{\partial x^2} +\frac{\varepsilon}{8} \frac{\partial^4\phi}{\partial x^4}+
\nonumber \\
&+& g \phi  \int_{-\infty}^{+\infty} G(x-x') |\phi (x')|^2 dx'
\end{eqnarray}
with $\mu>0$ being the soliton eigenvalue. Our aim next is to understand how the localization length of the bound states depends on the nonlinearity strength. In doing so, we shall consider soliton solutions corresponding to fixed initial power $P_0$, i.e.,
\begin{eqnarray}
\label{fixed_power}
\int |\phi|^2dx = P_0\;.
\end{eqnarray}
\subsection{Maximally localized nonlinear modes }
Solutions to Eq.(\ref{nls2}), in the form of a localized non\-li\-near waves, can be obtained by the spectral renormalization method \cite{Ablowitz:05}. To do so we define the renormalized complex wave function
\begin{equation}
\label{u}
\phi (x) = Ru(x)\;,
\end{equation}
where, in general, $R$ is a complex scalar, different from zero. Substituting (\ref{u}) into (\ref{nls2}) and (\ref{fixed_power}) gives expressions for both the soliton eigenvalue and the renor\-ma\-li\-zation factor
\begin{equation}
\label{mu}
\mu = \frac{|R|^2 E_{{\rm non}}(u) - E_k(u)}{ N(u) } \;.
\end{equation}
\begin{eqnarray}
\label{fixed_power2}
|R|^2 = \frac{P_0}{\int |\phi|^2dx}\;,
\end{eqnarray}
where we defined the ``kinetic", interaction energy and the power respectively:
\begin{equation}
\label{KE}
E_k(u) \equiv \frac{1}{2} \int |u_x|^2 dx +  \frac{\varepsilon}{8} \int |u_{xx}|^2 dx \;,
\end{equation}
\begin{equation}
\label{nonloc-int-E}
E_{{\rm non}}(u) \equiv  g \int \int G(x-x')  |u(x)|^2  |u (x')|^2 dx' dx
\end{equation}
\begin{equation}
\label{N}
N(u) \equiv \int |u|^2 dx\;.
\end{equation}
Using the one-dimensional Fourier transform defined in Eq. (\ref{FT-1}), we obtain
\begin{equation}
\label{iteration-1}
\hat{u} = \frac{ g |R|^2 \hat{u}*(\hat{G}\widehat{|u|^2}) }{\mu -k^2/2 - k^4 \varepsilon/8}\;,
\end{equation}
where
\begin{eqnarray}
\label{K-hat}
\hat{G}(k)=\sqrt{\frac{1}{2\pi}}\frac{1}{1+\sigma^2 k^2}\;.
\end{eqnarray}
Equation (\ref{iteration-1}) is a fixed point equation for $\hat{u}$ which can be solved by a direct fixed point iteration
\begin{equation}
\label{iteration-3}
\hat{u}_{n+1} = Q(\hat{u}_n,\mu_n, |R_n|^2)\;,
\end{equation}
where $|R_n|^2\equiv |R(u_n)|^2$ and  
\begin{equation}
\label{Q}
Q(\hat{u},\mu,|R|^2) = \frac{g |R|^2 \hat{u}*(\hat{G}\widehat{|u|^2}) }{\mu + k^2/2 + k^4\varepsilon/8} .
\nonumber
\end{equation}
\begin{figure*}[t!]
	\centering
		\includegraphics[scale=0.5]{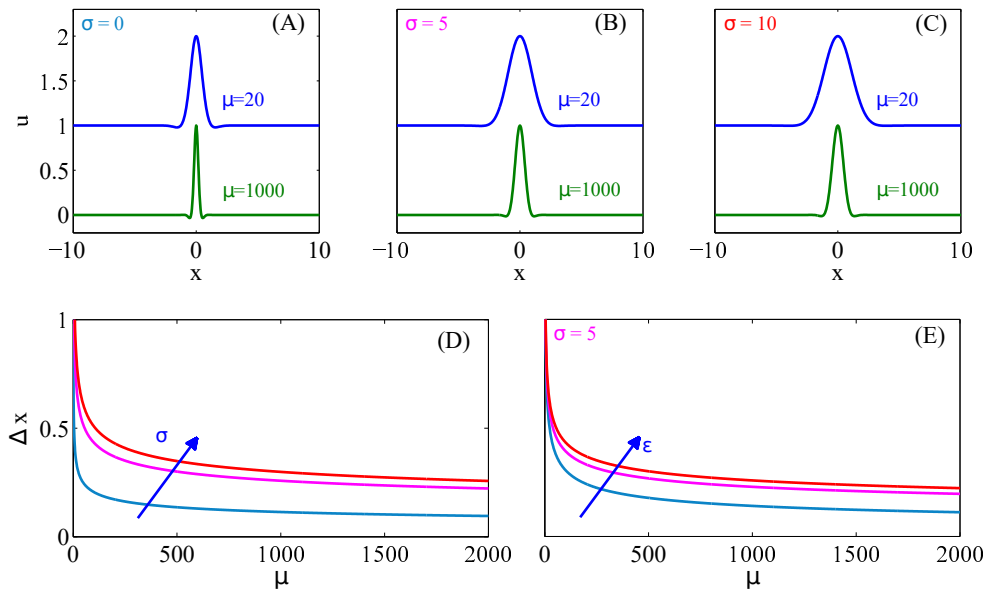}
	\centering
	\caption{(A) Normalized fields $u$ after Eq. (\ref{nls2}) and (\ref{u}), for high and low values of the eigenvalues $\mu$ ($\mu=20$ or $1000$) in the case of local nonlinearity. The curve at low $\mu$ has been shifted on the vertical axes to allow a clearer view of the lobes. (B) and (C) as in (A) for degree of nonlocality $\sigma=5$, $10$ respectively. (D) Behavior of the width $\Delta x$ as a function of the eigenvalue $\mu$ for different nonlocality $\sigma$ ($\sigma=0,\, 5,\,10$); (E) as in (D) but at different values of the degree of nonparaxiality $(\varepsilon=0,\,0.5,\,1)$ at fixed nonlocality $\sigma=5$. We used large values for $\varepsilon$ to emphasize the differences among the lines.} 
\label{fig:2}
\end{figure*}
In figure \ref{fig:2} we show the bound states calculated with the spectral renormalization method. At fixed nonlinearity $g$, we study the soliton width by varying the eigenvalue 
$\mu$, that is equivalent to varying the solitary wave. We observe that at high $\mu$ the wave profile develops lateral lobes (bottom curve in panels A,B and C of Fig. \ref{fig:2}) as expected for the maximally localized state (see Fig.\ref{fig:psix}A). These lobes becomes smoother as increasing the degree of nonlocality $\sigma$. In panels D and E of Fig. \ref{fig:2} we report the behavior of the soliton width as a function of power. It results that $\Delta x$ increases for higher values of $\sigma$. The same result is obtained varying the degree of nonparaxiality $\varepsilon$. It is worthwhile to notice that for increasing $\mu$ the width tends to saturate to a lower value, i.e., to the maximal localization.   
\begin{figure*}[t]
	\centering
		\includegraphics[scale=0.5]{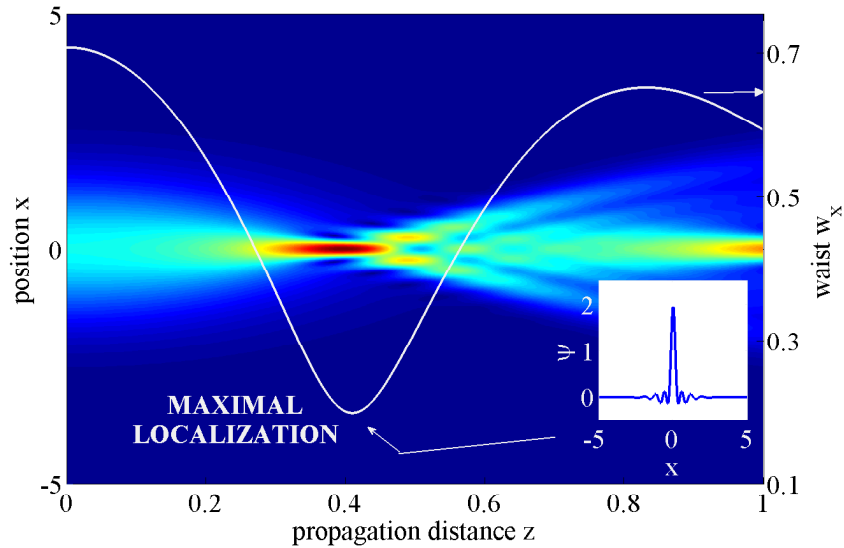}
	\centering
	\caption{Simulation of a Gaussian beam evolving according to Eq. (\ref{nls2}), for $g=1$, $\varepsilon=10^{-5}$ and $\mu=10^4$ . The superimposed white line shows the waist $w_x$ versus the propagation direction $z$. The inset shows the field profile at the point of maximal localization ($z\simeq0.4$).}
\label{fig:5}
\end{figure*}
\subsection{Excitation of maximally localized states}
In order to provide a further evidence that non\-li\-nea\-rity forces the system towards maximal localization, we numerically solve Eq. (\ref{nls1}) with kernel (\ref{kernel}). The initial beam profile is a Gaussian beam [see Eq. (\ref{gaussian_beam})]. Fi\-gu\-re \ref{fig:5} shows that the beam focuses upon propagation and its waist $w_x$ presents a minimum (maximal localization) during propagation. As the inset shows, the field at the maximal localization displays the characteristic lateral lobes, with a remarkable resemblance with Fig. \ref{fig:psix}A.\\
Albeit, these results confirm the onset of maximal loca\-li\-za\-tion, we remark that when the beam waist is comparable with the minimal length the first order perturbation theory used in Eq. (\ref{NORMALIZED}) looses validity. This calls for more advanced theoretical methods that will be reported in future works.

\section{Conclusions}
We have reported on the implementation of the quantum gravity generalized uncertainty principle in the nonlinear Schr\"{o}dinger equation and provided an analogue to study QG effects thanks to optical propagation. We considered the simplest form of the theory based on a generalized linear Schr\"{o}dinger equation with higher order dispersion/diffraction. This equation describes the propagation of ultra-short pulses in fibers or one-dimensional sub-paraxial focused beams. We have discussed the way a generalized uncertainty principle enters in the description of possible states. We have analyzed the resulting maximally localized states and shown the way they can be excited in nonlinear propagation. Our goal was to demonstrate that ideas from quantum gravity have relevance in optics and photonics including the nonlinear regime. This analysis might be extended in several directions such as retaining higher order dispersion and calculating the shortest pulse that can propagate in a fiber at any dispersion order. Another possibility might be designing spatially modulated beams in order to ultra-focus beyond the limits imposed by standard numerical aperture.\\Developments also include novel classes of nonlinear waves in the spatio-temporal domain. Furthermore our results show that photonics can be an important framework to realize analogues or models of Quantum Gravity theories. \cite{Faccio2012bh,Barcelo2003,Longhi2011,Michinel_16}\\

We acknowledge fruitful discussions with D. Faccio, R. Boyd, F. Biancalana and E. Wright. 
This publication was made possible through the support of a grant from the John Templeton Foundation (58277). The opinions expressed in this publication are those of the author and do not necessarily reflect the views of the John Templeton Foundation. We also acknowledge support by the European Research Council Grant ERC-POC-2014 Vanguard (664782).


\end{document}